\documentclass[pre,reprint,superscriptaddress]{revtex4-1}
\usepackage{graphicx}
\usepackage{epstopdf}
\usepackage{amssymb,amsfonts,amsmath}
\usepackage{color}
\usepackage{float}
\usepackage{gensymb}
\usepackage{dcolumn}
\usepackage{bm}
\usepackage{wrapfig} 
\usepackage{xcolor}
\usepackage{mdframed}
\usepackage[many]{tcolorbox}
\usepackage{lmodern}
\usepackage{lipsum}
\usepackage{amsmath, nccmath}
\usepackage{multirow}
\usepackage{tabu}
\usepackage{float}
\usepackage[caption=false]{subfig}
\usepackage{capt-of}
\newcommand{\RN}[1]{\textup{\uppercase\expandafter{\romannumeral#1}}}
\usepackage{algorithm}
\usepackage[noend]{algpseudocode}
\usepackage{esvect}
\usepackage{mathtools}
\usepackage{siunitx}

\usepackage[colorlinks=true,linkcolor=blue,urlcolor=blue,citecolor=blue]{hyperref}

\usepackage{soul}
\begin{document}
\title{Elasticity of a DNA chain dotted with bubbles under force }
\author{Debjyoti Majumdar}
\email{debjyoti@iopb.res.in}
\affiliation{Institute of Physics, Bhubaneswar, Odisha 751005, India}
\affiliation{Homi Bhabha National Institute, Training School Complex, Anushakti Nagar, Mumbai 400094, India}
\date{\today}
\begin{abstract}
The flexibility and the extension along the direction of the force are shown to be related to the bubble number fluctuation and the average number of bubbles respectively, when the strands of the DNA are subjected to a force along the same direction, here we call a stretching force. The force-temperature phase diagram shows the existence of a tricritical point (TCP), where the  first-order force induced zipping transition becomes continuous. On the other hand, when the forces are being applied in opposite directions, here we call an unzipping force, the transition remains first-order, with the possibility of vanishing of the low-temperature re-entrant phase for a semiflexible DNA. Moreover, we found that the bulk elasticity changes only if an external force penetrates the bound phase and affect the bubble states. 
\end{abstract}
\maketitle
\section{Introduction} 
Perturbing a polymeric system and looking at its response paves the way for us to probe its thermodynamical and structural properties in different phases. Pulling a stiff linear object subjected to thermal fluctuations is in itself an important class of problem  \cite{odijk,oosawa,morrison}.  Double-stranded DNA (dsDNA) is one such polymeric system which undergoes a thermal melting transition due to the breaking of the hydrogen bonds holding the base-pairs together, which is called DNA melting. This opening up of the DNA into two single strands is the first step towards fundamental biological processes such as DNA replication, RNA transcription etc. and is often initiated by enzymes like helicase, polymerase etc. which exerts force to open up specific sections of the DNA \cite{watson}. Consequently, the functionality of the DNA, which depends upon its bulk properties such as elasticity, length etc. might get altered under the action of these regulatory forces. 
 
With the advancement of technology now it is possible to manipulate forces  at the microscopic level using optical, magnetic tweezers or atomic force microscopic techniques \cite{smith,wenner}.  This is where mechanical ways like unzipping and stretching become important. For example, a DNA can be mechanically  unzipped \textit{in vitro} by applying an external force to separate the two strands apart, which in turn can provide us with information regarding the hydrogen bonds holding the strands together along the base pairs, thus revealing the heterogeneous nature of the base-pair sequence \cite{smb1,rief,orlandini1,boland,chrisey,roulet,bockelmann0,bockelmann1,dani0,granek}. Since force-induced melting transition is isothermal in nature, one can avoid the poorly characterized thermal contributions to the transition entropy and enthalpy, thus giving an extra advantage over thermal melting transitions \cite{li}. Different types of phase transitions have been observed \cite{kapri2, kgb}. These melting transitions are associated with a change in the elastic property due to a change in the topology of the system e.g. elasticity of the DNA changes when the ribbon picture is lost \cite{brunet,dm,comm6}. Thus, it is important to understand how these forces might affect the DNA elastic properties under a change in the state of the system. 
 
 Experimentally, the elastic properties of a DNA is studied from the force-extension curves obtained from single molecular experiments \cite{marko,smith}. These curves are then fitted with various theoretical models.  In the single chain limit the energy associated with the conformational fluctuations of the DNA can be modelled using the linear elasticity of a thin rod \textit{\`a} \textit{la} the worm-like chain (WLC) model, described by the following hamiltonian 
\begin{equation}
    \mathcal{H}_{\rm wlc}=\frac{1}{2}\kappa \int_{0}^{N} \left( \frac{\partial^2 
    \bf{r}(s)}{\partial s^2}\right)^2ds,
\label{eq:1}
\end{equation}
where $\kappa$($\equiv k_BT l_p$) is the bending elastic modulus and $l_p$ is the characteristic length scale, called the persistence length, over which the rodlike behavior is maintained, $N$ is the length of the chain, $s$ is the arc length along the semiflexible chain, $k_B$ is the Boltzmann constant and $T$ is the temperature. The WLC model predicts that in the presence of a force ${\bf f}$ the fractional extension along the direction of force $\zeta=1-(z/N)$, where $z$ is the extension along the direction of force ($\hat{\bf z})$, of a semiflexible chain, goes to zero as $\zeta\sim f^{-1/2}$ while from the FJC model the extension scales with force as $\zeta\sim f^{-1}$, in the large force limit \cite{marko,smbgm}. The only parameter in the description of these models is the persistence length ($l_p$). These predictions do not take into account the presence of thermally denatured (broken hydrogen bonds) local regions, known as thermal bubbles, which act as local hinges for the DNA to make bends and gain flexibility. This should lead to an effective renormalized elasticity $\kappa$ (or equivalently $l_p$) of the whole DNA \cite{amnuanpol}. Recently, it has been shown, that near the melting transition in the zero-force limit ($f=0$), the elastic modulus might be more meaningful than the notion of a persistence length ($l_p$) \cite{dm}. Since the usual definition of the persistence length from the tangent-tangent correlation may not be meaningful, the persistence length loses its usual significance. In such a situation it is reasonable that the validity of the WLC model is questionable . Besides providing flexibility, these bubbles are also associated with important biological functions. In recent times, many investigations have been performed regarding the various aspects of these bubbles, which includes {change in rigidity \cite{dm},} breathing dynamics \cite{altan,ambjornsson,alex,sicard}, in hysteresis \cite{rkapri,mishra}, for DNA under topological constraint \cite{son,king,jeon,strick,adamcik}, for semiflexible DNA \cite{sung}, etc.

Our aim in this paper is to study the elastic properties of a flexible and semiflexible DNA under unzipping and stretching type forces in the presence of thermally melted regions or bubbles using Monte-Carlo simulations on a cubic lattice (d=$3$)  while revisiting some earlier results from the perspective of our model. Semiflexibility is introduced in the bends of the ds segments, while ss segments are exempted from any such energy costs; see section \RN{2} for details. In reality the ds segments are much more stiffer ($l_p=150$ bp or $\approx 50$ nm) than the single-stranded (ss) or unbounded segments ($l_p=4$ bp or $\approx 2.5$ nm) \cite{hagerman,rechendorff,abels}. We also investigate how the bubble statistics and hence the nature of the thermal melting transition as shown by this particular model, gets modified under various forcing conditions. Our focus will be on the regime of intermediate forces, where a coarse-grained 
picture of the DNA is valid and microscopic details such as the local structure, bond length, bond angles, torsional potentials etc. remains irrelevant. {High forces on the other hand shows interesting phenomena \cite{zhang}. Although such regime is out of purview of our study.}

\begin{figure}[t]
    \centering
    \includegraphics{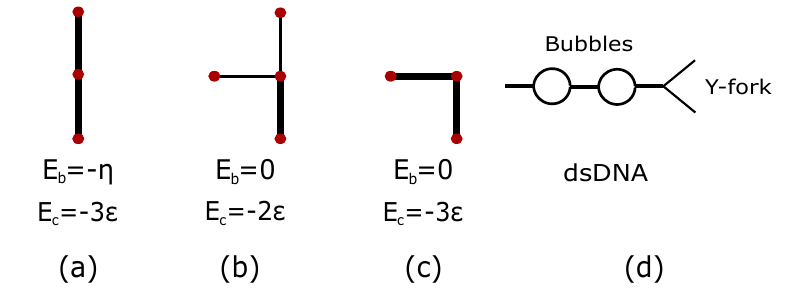}
    \caption{(Color online) Possible configurations for a two step walk on a plane for three consecutive monomers ({red dots}) and energies associated with contact ($E_c$) and bending ($E_b$) according to Eq.~\ref{fig:2}. (a) Three contacts resulting into contact energy $E_c=-3\epsilon$ and shifted bending energy $E_b=-\eta$, (b) opening of a Y-fork with two contacts $E_c=-2\epsilon$ and (c) same as (a) but with a bend, costing a bending energy (shifted) $E_b=0$.  {The red dots also represent lattice sites.} (d) Identifying bubbles and the Y-fork in a dsDNA.}
    \label{fig:1}
\end{figure}

This paper is organized in the following manner. In Section $\RN{2}$ we have described our model and discussed the observables required to study the elastic properties. In Section $\RN{3}$ we have discussed the algorithm for simulating the dsDNA on the cubic lattice and for applying an external force at the endpoints.  Section $\RN{4}$ and $\RN{5}$ deals with the elastic response of the DNA under stretching and unzipping forces respectively. And section $\RN{6}$ 
concludes the paper.

\begin{figure}[t]
\centering
\includegraphics{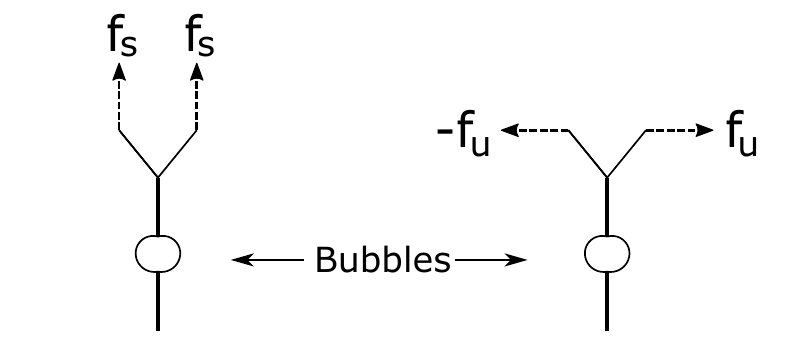}
\caption{Schematic representation of a stretching force ($f_s$) and an unzipping force ($f_u$) on a dsDNA consisting of a {\rm Y}-forklike region at the end and a bubble embedded between closed segments. The two diagrams represent a DNA under stretching (left) and unzipping (right) forces respectively.}
\label{fig:2}
\end{figure}


\section{DNA Model and Qualitative Description}
Our minimal model for the DNA consists of two linear polymer chains on a cubic lattice \cite{causo}, which are self as well as mutually avoiding with the exception that they can form energetically favourable contacts with energy  $E_c=-\epsilon(\epsilon>0)$ only at the same monomer position along the chain. Through out the simulation we have chosen $\epsilon=1$.  One end of the DNA is fixed while the other end is free to wander. We consider two different cases which we call the flexible and the semiflexible model. In the flexible model, we consider two self and mutually avoiding walks with complete flexibility even in the bound state.  While in the semiflexible model we associate an energy with bending of the ds or bound segments. The energy for bending of the ds segments is given by 
\begin{equation}
    E_{\rm b}=-\eta \cos{\theta}, 
    \label{eq:2}
\end{equation}
where $\eta(>0)$ is the bending energy constant.  An increased $\eta$ means a stiffer chain. Note that the Boltzmann weight $\exp{(-E_{\rm b}/k_BT)}$ for a straight move $(\theta=0 \degree)$ remains higher than that for a bend $(\theta=90 \degree)$; see Fig.~\ref{fig:1}(a) - (c). 

To investigate the elastic properties we apply a space independent constant external force $\textbf{f}=f\hat{\textbf{z}}$  {in the direction $\hat{\textbf{z}}$} (fixed force ensemble) at the free endpoints $\textbf{r}_i(N)$ of each strand $i=1,2$, while the other end remains fixed at the origin. If the forces on the two strands are in the same direction they are said to be the stretching force, denoted by $f_s$, while an unzipping force $f_u$ pull the two strands of the DNA in the opposite directions; see Fig.~\ref{fig:2}. This minimal model allows for the formation of bubbles which, as we will see, plays an important role in determining the elastic response of the system under a stretching force. Both the models show $(i)$ a zero-force melting temperature, which we call the thermal melting point, generically defined as $T_m$ \cite{dm}, $(ii)$ a stretching induced zipping transition from an unbound state to a stable bound state beyond a critical stretching force $f_{sc}$ for any $T>T_m$ and $(iii)$ an unzipping phase transition beyond a critical force $f_{uc}$ at any finite temperature below $T_m$.

The {canonical} partition function of a $N$ length DNA, where $N$ is the number of bonds, in the presence of a stretching force ${\bf f}_s$ can be written as 

\begin{equation}
\mathcal{Z}({\bf f}_s,\beta)=\sum_{{\bf r}_1,{\bf r}_2}\mathcal{C}_N(
{\bf r}_1,{\bf r}_2) e^{\beta{\bf f}_s.{\bf R}},
\label{eq:3}
\end{equation}

where $\mathcal{C}_N({\bf r}_1,{\bf r}_2)$ is the zero-force partition function of the $N$ length DNA with end position vectors ${\bf r}_1$ and ${\bf r}_2$, ${\bf R}(N)={\bf r}_1(N)+{\bf r}_2(N)$ is the vectorial position of the center-of-mass (c.m.) of the end points, and the sum is carried out over all possible values of ${\bf r}_1$ and ${\bf r}_2$ and $\beta(=1/k_BT)$ is the inverse temperature. We set the Boltzmann constant $k_B=1$ through out our work. The elastic response of the DNA under an external force is quantified using the tensorial quantity $\chi_{ij}$, defined in  the following way
		
\begin{equation}
    {\zeta_{\rm cm}}_i=\frac{1}{\beta}\frac{\partial\ln{\mathcal{Z}}}{\partial {f_s}_ i}~~\text{and}~~\chi_{ij}=\frac{\partial {\zeta_{\rm cm}}_i}{\partial {f_s}_j},
    \label{eq:4}
\end{equation}

where $\zeta_{{\rm cm}_i}$ is the average extension of the i-th component of the c.m. chain and the average c.m. position is written as 

\begin{equation}
 {\bm \zeta}_{\rm cm}=\frac{\sum_{{\bf r}_1,{\bf r}_2}\mathcal{C}_N({\bf r}_1,{\bf r}_2) e^{\beta{\bf f}_s.{\bf R}}{\bf R}}{\sum_{{\bf r}_1,{\bf r}_2}\mathcal{C}_N({\bf r}_1,{\bf r}_2) e^{\beta{\bf f}_s.{\bf R}}}.
 \label{eq:5}
\end{equation}
 
Then, in the zero-force limit, where the anisotropy in shape is isotropic in all directions \cite{comm3}, we can relate the {elastic response function (e.r.f)} $\kappa_{\rm cm}$ to the fluctuations in the vectorial position of the c.m. position of the end points as \cite{dm}

\begin{eqnarray}
\bar{\kappa}_{\rm cm}&\equiv&\frac{1}{\beta}\kappa_{\rm cm}= \frac{1}{\beta} {\rm Tr}\left[ {\bm \chi} \right]=\left<\textbf{R}(N)^2\right>-\left<\textbf{R}(N)\right>^2\\
    &=&2\langle
    {\bf r}_1(N)^2\rangle_c \left(1+\frac{\langle  {\bf r}_1(N)\cdot{\bf
    r}_2(N)\rangle_c}{\langle  {\bf r}_1(N)^2\rangle_c}\right ),
    \label{eq:7}
 \end{eqnarray}

where {the subscript $c$ refers to the second cumulant} and the factor of $2$ comes from the symmetry between the two strands. {Notice that when ${\bf r}_1$ and ${\bf r}_2$ are uncorrelated $\bar{\kappa}_{\rm cm}$ is the sum of the elastic response of the individual strands.} For Gaussian chains {$\bar{\kappa}_{\rm cm}/\langle {\bf r}_1(N)^2\rangle_c$} is exactly $2$ in the unbound phase, and for chains with excluded volume interaction this is {slightly} greater than $2$ {due to interstrand correlation} \cite{degennes}. The isotropy breaks down in the presence of any external force, consequently other off-diagonal terms in the ${\bm \chi}$ tensor becomes important e.g.,

\begin{equation}
\chi_{xy} = 2\left[ \langle r_{1x} r_{1y}\rangle_{c} + \langle r_{1x} r_{2y} \rangle_{c}\right].
\label{eq:8}
\end{equation}

 Although, in this study we will focus on the isotropic part only, since, this would facilitate comparison with the zero-force scenario \cite{dm} and a further investigation of the other off-diagonal terms is left for a future study. According to the definition of the {e.r.f} as in Eq.~\ref{eq:4} and \ref{eq:7}, a higher value of $\bar{\kappa}_{\rm cm}$ denotes an increased flexibility under an applied force. Another quantity similar to $\bar{\kappa}_{\rm cm}$ comes from the relative (rel) chain for forces along the opposite directions  and is obtained from Eq.~\ref{eq:7} by replacing the positive sign with a negative as

\begin{eqnarray}
\bar{\kappa}_{\rm rel}=2\langle
    {\bf r}_1(N)^2\rangle_c \left(1-\frac{\langle  {\bf r}_1(N)\cdot{\bf
    r}_2(N)\rangle_c}{\langle  {\bf r}_1(N)^2\rangle_c}\right ).
    \label{eq:9}
\end{eqnarray}
 
Interestingly, the c.m. chain is not a conventional polymer except in special situations e.g. $T\rightarrow 0$ with no bubbles. This makes the dsDNA rigidity problem different from a simple minded single polymer problem. Moreover, when bubbles coexist with semiflexible bounded segments, the c.m. chain {behaves like a} multiblock copolymer constructed from hard rods (semiflexible ds segments) and flexible chains (bubbles).

The nature of the {e.r.f} obtained from $\bar{\kappa}_{\rm cm}$ and $\bar{\kappa}_{\rm rel}$, as we will see, depends on how the forces are applied at the two endpoints. This is similar to the direction-dependent elastic response on pulling a single strand polymer (protein) from collapsed or globule state to an extended state when the model is inherently anisotropic in shape and conformation \cite{kumargiri1}. Anisotropy in our model is introduced by the application of an external force. Therefore, while one of the force introduces anisotropy in shape, varying the direction of the other force leads to a different elastic response.

Note that, while the bound state of the flexible model ($\eta=0$) the ds remains as flexible as the ss, as a result there is only emergent entropic elasticity. On the  other hand, in the semiflexible model ($\eta \neq 0$) the bound state has an intrinsic rigidity  towards bending and the only way to gain flexibility is through the formation of locally melted bubbles. The model we considered here is that of a torsionally unconstrained DNA where the helical topology is disregarded. Although, the semiflexibility due to the stacking of the base pairs in the helix is effectively included as the semiflexibility of the ds bounded segments.

\section{Simulation Algorithm}      
 For simulation we have used the $\textsc{PERM}$ (Pruned and Enriched Rosenbluth Method) algorithm, which sample equilibrium configurations of long chains efficiently through successive cloning and pruning approach controlled by a predefined threshold \cite{causo,prellberg}.  Both the chains take new steps at the same instance, the choices of which are given by the joint possibilities of atmospheres of both the chains. The monomers are added to a chain successively, one after another, following the $\textsc{Rosenbluth-Rosenbluth}~(\textsc{RR})$ method \cite{rosenbluth}.  At each step, the local partition function is calculated by estimating all the possible configurations with proper Boltzmann weights. Since for self-avoiding chains the atmosphere seen by the open end of the two chains can be different we need to consider the combined atmospheres of the two chains \cite{comm1}.  Since overlap is allowed only at the same monomer position of the two strands, the total atmosphere would simply then be the multiplication of the individual atmospheres of the two strands, i.e. the total atmosphere at $n$th step for two self and mutually-avoiding chains would simply be $atmos=atmos_{1}\times atmos_{2}$, where $atmos_{1}$ and $atmos_{2}$ refers to the atmospheres of strand $1$ and $2$ respectively. Then for two interacting walks, the first step for each walk has 6 different possibilities, with a total of 36 combined possibilities to step into, among which $6$ ways of taking steps together and form a bond. Thus, the local partition function becomes $\mathcal{Z}_{\rm local}=w_1=30+6\exp{(\epsilon/T)}$, where $\exp{(\epsilon/T)}$ is the Boltzmann factor for making a contact and $\mathcal{Z}_{\rm local}$ also serves as the weight of that particular step $w_{1}$.  Then the weight of a configuration of length $N$ is $W_{\rm N}=\prod_{i=1}^{\rm N}w_{i}$, the successive multiplication of the weights of the previous steps. Likewise, the weight of the second step  {provided that the first step is bound} would be $w_2=4exp(\epsilon/T) + exp(\epsilon/T)exp(\eta/T) + 20$. The external force is introduced similarly.  Assuming that the strands are stretched along $z$-direction by a force $\textbf{f}=f\hat{\bf z}$ ($f$ is the magnitude of force and $\hat{\bf z}$ is the unit vector along the $z$-direction) a weight factor $b=\exp{(\Delta zaf/T)}$ is introduced in calculating the weight $w_n$ at the $n$th step for both the chains
\begin{multline}
   w_n = \sum_{atmos}\exp{(-E_c/T)} \exp{(-E_b/T)} \exp{(\Delta z_1 af_{1}/T)}\\
    \times \exp{(\Delta z_2 af_{2}/T)}
 \label{eq:10}
\end{multline}

where $f_1$ and $f_2$ are the forces at the endpoints of chain 1 and 2 respectively and the sum is over all free directions ($atmos$) with $E_c$ and $E_b$, the contact energy and the bending energy respectively. $\Delta z_{1,2}=\pm 1$ for step along or opposite to the direction of the force and $0$ otherwise. $a(=1)$ is the step length. Enrichment and pruning are performed at each step depending on whether the $ratio=\frac{W_{n}}{\mathcal{Z}_{n}}$ is greater or smaller than $1$.  In our simulation averages are taken over $10^{7}$ tours and error bar for the fluctuating quantities are estimated on the fly. {For a discussion on error calculation see \cite{dm}}. The averaging is not as simple as in the case of the $\textsc{RR}$ method, where a tour is just a single chain.  For $\textsc{PERM}$ a tour is a set of chains with a rooted tree topology, where new branches are added through cloning and moves are performed only along the branches of the tree to ensure detailed balance. {To check self avoidance, we implement two different methods. One with a virtual box where lattice sites are indexed 0 or 1 if empty or occupied respectively. Although, time complexity is $\mathcal{O}(1)$, memory required is large thus limiting the maximum length that can be achieved.  In the second method we used a tree based search algorithm known as $\textsc{AVL}$ tree. Here the time complexity is $\mathcal{O}\left(\log n\right)$ in the worst case, where $n$ is the number of nodes.}


\begin{figure}[t]
\centering
\includegraphics[width=\linewidth]{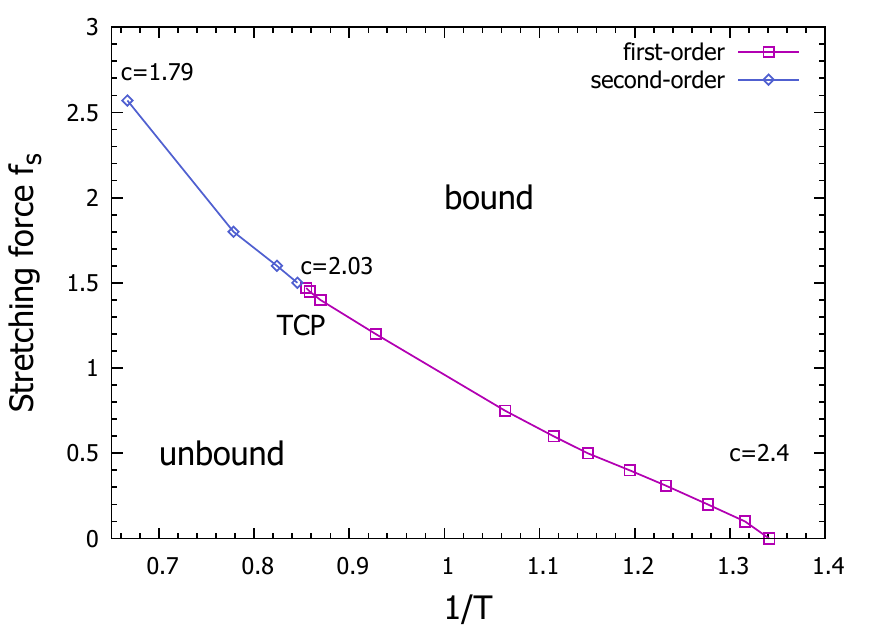}
\caption{(Color online) Force-temperature phase diagram for a flexible DNA ($\eta=0$) under a stretching force ($f_s$). The red and blue colored points represents the first- and second-order phase boundaries respectively which meets at a tricritical point (TCP). {We have shown the bubble-size-exponent $c$ along the transition line.}}
\label{fig:3}
\end{figure}

\vfill

\section{Elastic Properties under a Stretching Force}    
\begin{figure}
\centering
\hspace*{-8cm}(a)\\\includegraphics{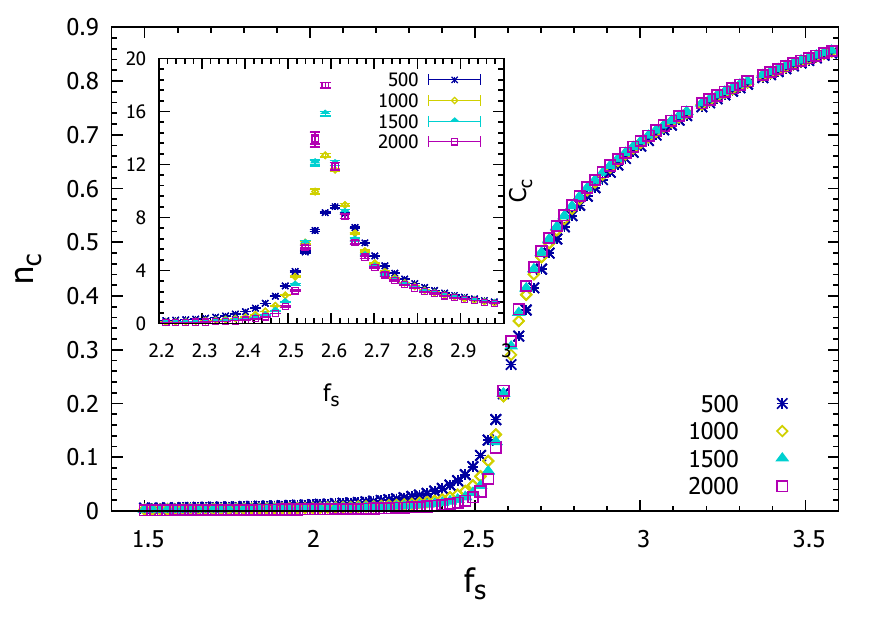} 
\hspace*{-8cm}(b)\\\includegraphics{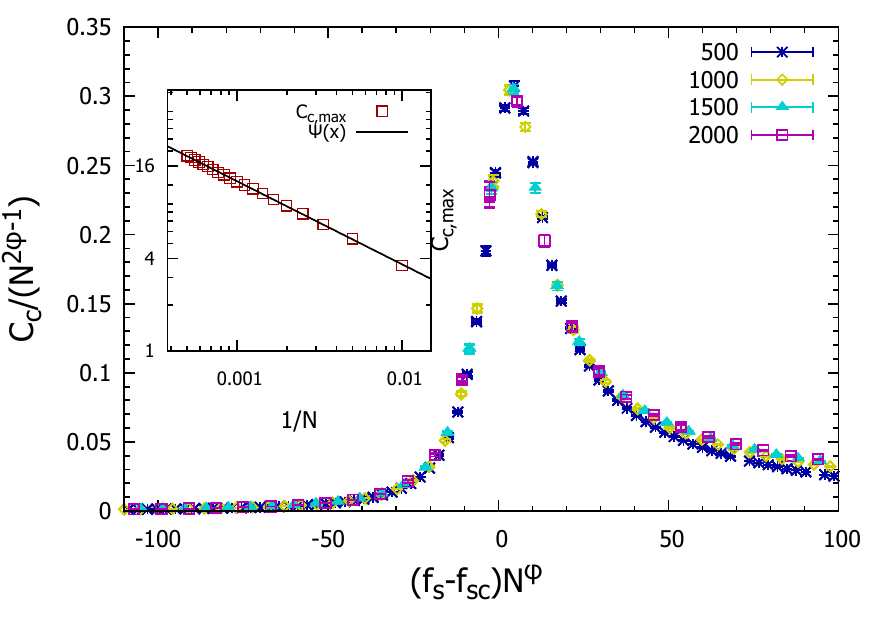}
\hspace*{-8cm}(c)\\\includegraphics{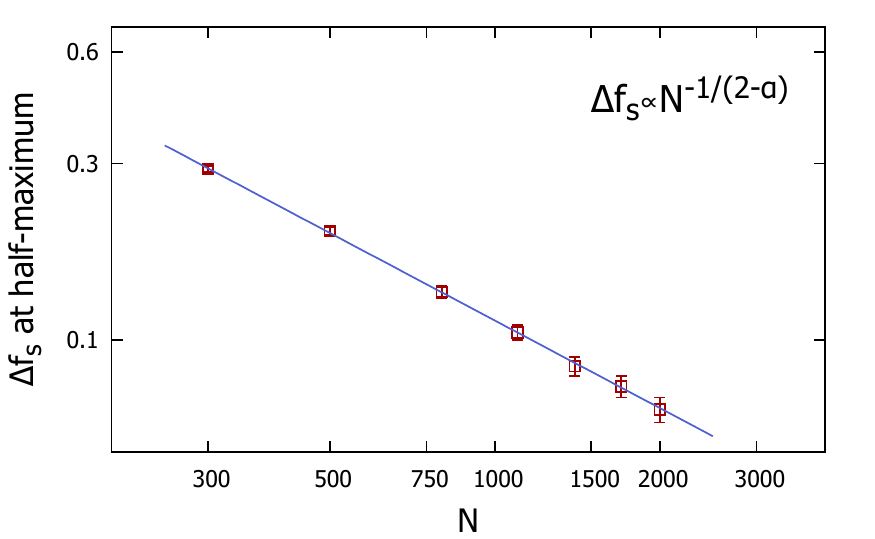}
\caption{(Color online) {Order parameter, its fluctuation, data collapse and smearing exponent. (a) Number of base-pair contacts per monomer $(n_c)$, for a flexible DNA i.e $\eta=0$ under a stretching force $(f_s)$ at temperature $T=1.5$($>T_m$). (Inset) Contact number fluctuation per monomer $(C_c)$. (b) Data collapse of $C_c$ curves according to Eq.~\ref{eq:12} using $\phi=0.77$ and $f_{sc}=2.57$. (Inset) Log-log plot for scaling of the contact fluctuation peaks $C_{c,max}$ with the system size $N$.} $\Psi(x)\sim x^{-0.54}$ is a fit to the data points resulting in $\phi=0.77\pm0.001$. (c) Log-log plot of $\Delta f_s$ with $N$. Smearing exponent quantifying the finite-size rounding of the $(C_c)$ curves (Fig.~\ref{fig:4}(a) inset) at $T=1.5$ and $\eta=0$. We obtain $\alpha=0.72 \pm 0.009$ from fitting the data points using Eq.~\ref{eq:13}} \cite{imry}.
    \label{fig:4}
\end{figure}

\textit{The zipping transition:} If the forces applied at the end of the two wandering strands of the DNA are towards the same direction, then the forces are said to be stretching in nature; see Fig.~\ref{fig:2}.  A model similar to ours, but over a directed lattice, was considered here \cite{pal,marenduzzo}, where a continuous transition was observed. {For directed lattice, the phase transition behavior in d+1 dimension is the same as that of Gaussian chains in d dimension.} Here, we consider self-avoiding interaction among all segments of the chain and same force upon both the strands. Although, the case of unequal forces could be equally interesting, as this would raise the question regarding the possibility of a bound phase {at} small forces, {for a DNA interacting only at the complementary sites.} The simultaneous stretching of both the strands along the same direction has a stabilising effect on the bound state of the DNA. Under stretching, the strands tend to come together and form contacts which carry the system from an unbound state to a stable bound state. This happens above a critical stretching force ($f_{sc}$) which depends upon the temperature $(T)$ and the rigidity $(\eta)$ of the system. The critical stretching force required to zip the DNA increases for a DNA at an elevated temperature or with higher flexibility. In other words, the semiflexible chain tends to cooperate with the stretching force.  While the thermal melting transition $(f_s=0)$ with excluded volume interaction is first-order \cite{dm,causo}, under a stretching force the DNA undergoes a continuous zipping transition even with excluded volume interaction, for sufficiently strong  forces \cite{mdz}.  This is evident from the shift of the specific heat ($C_c$) peaks with the system size (Fig.~\ref{fig:4}(b) inset) or the bubble-size-exponent ($c$). This can be explained from the correlation of the fluctuations along the polymer chain, also known as the ``deflection length" $\lambda$, which in the presence of a strong force is given by \cite{hsu,odijk2,odijk3,livadaru}

\begin{equation}
    \lambda/l_p=(f_sl_p/k_BT)^{-1/2}.
    \label{eq:11}
\end{equation}

\begin{figure}
    \centering
     \hspace*{-8cm}(a)\\\includegraphics{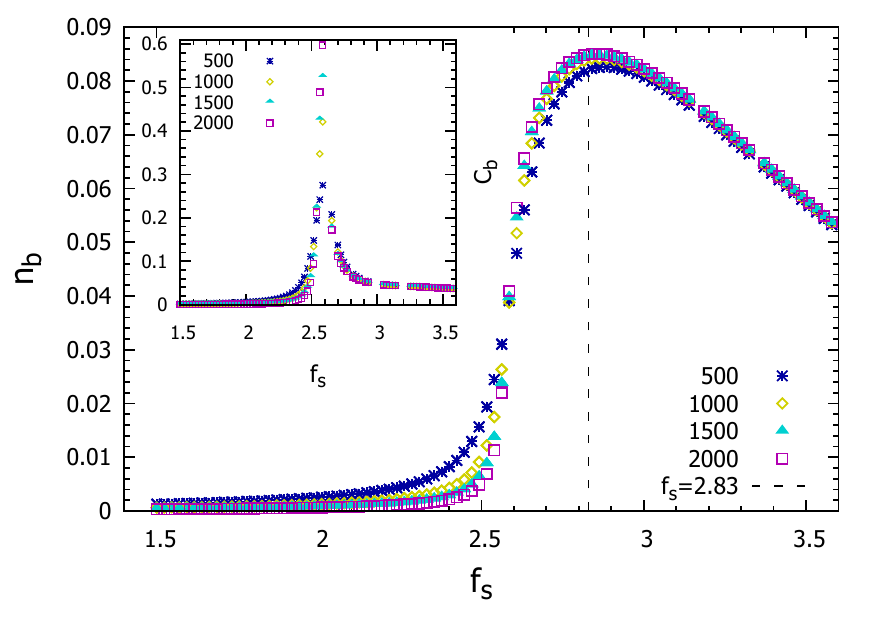}
     \hspace*{-8cm}(b)\\\includegraphics{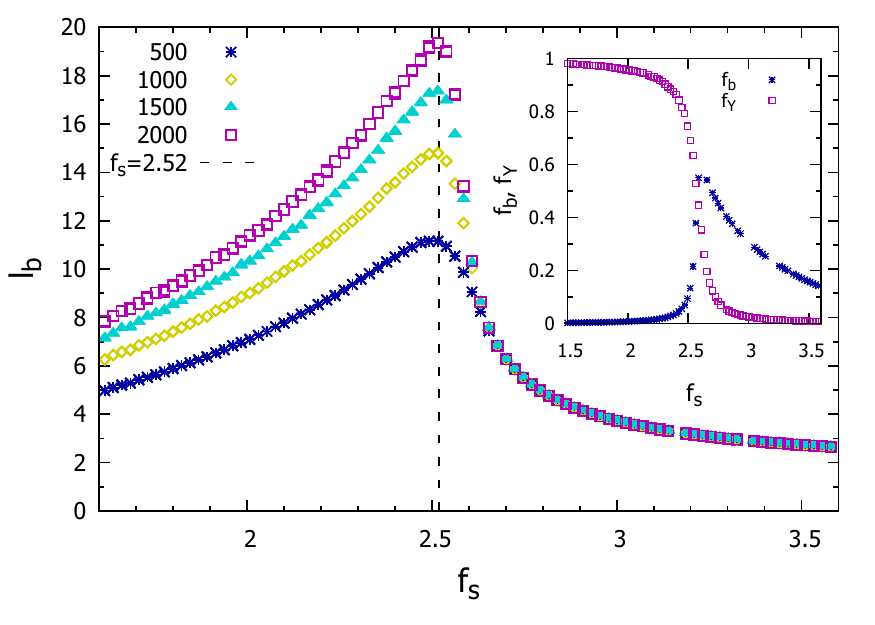}
     \hspace*{-8cm}(c)\\\includegraphics{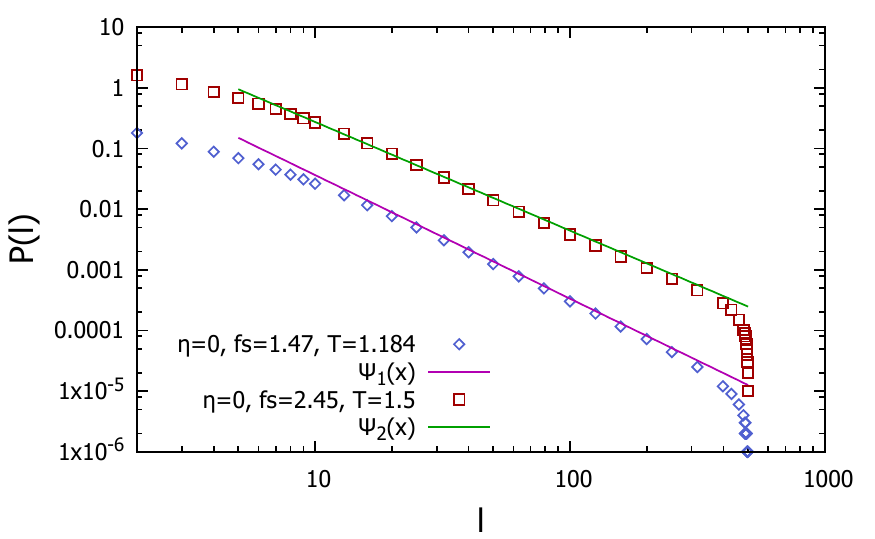}
    \caption{(Color online) Bubble {statistics}. (a) Average number of bubbles per monomer $(n_b)$ against the stretching force $(f_s)$ for a flexible DNA ($\eta=0$)  at $T=1.5(>T_m)$ for chain length up to $N=2000$. (Inset) Same as panel (a), but with bubble number fluctuation per monomer $(C_b)$ along the y-axis. (b) Average bubble length ($l_b$) along the chain for a flexible DNA $\eta=0$ at $T=1.5$, for chain lengths upto $N=2000$. Average taken over a chain and then over configurations. (Inset) Fraction of broken bonds that forms bubbles $(f_b)$ and Y-fork $(f_Y)$. (c) Bubble-size-distribution $P(l)$ near the tricritical point $f_s=1.47$ and $T=1.184$ {with $c=2.03\pm 0.001$}; see Fig.~\ref{fig:3}. {Also at $T=1.5$ and $f_s=2.45$ {with $c=1.79\pm 0.01$}. Both for} chain length $N=500$. $\Psi_1(x)\sim x^{-2.03}$ {and $\Psi_2(x)\sim x^{-1.79}$} is a fit to the data points along the middle of the distribution. {Data points for $\psi_2(x)$ are shifted by a factor of $10$.}}
\label{fig:5}
\end{figure}

 When $\lambda$ becomes $l_p$ the excluded volume becomes irrelevant and thus the DNA undergoes a continuous 
 renaturation transition. In simple words, the reduced interaction between the bubbles and the rest of the chain due to the presence of the external force results in this change of nature of the transition \cite{hanke, kafri, carlon}.  This hints at the possibility of the existence of a special point in the force-temperature phase diagram where the transition changes from a second-order to a first-order or vice-versa and is known as a tricritical point (TCP). The TCP {has been} determined by observing the value of the exponent $c$ {near the critical point}. For the exponent $c < 2$ the transition is regarded as second-order, while $c\geq 2$ the transition is first-order. {First-order thermal melting follows an exponent $c=2.4$} \cite{dm}. The strongest first-order transition {reported is for}  $c=3.2$ \cite{carlon}. We obtain $c$ directly from the bubble-size-distribution (Fig.~\ref{fig:5}(c)) for various $T$ near the transition point $f_{sc}$, which in turn is estimated from the specific heat curves. {The TCP is obtained at $f_s=1.47$ and $T=1.184$ with $c=2.03\pm 0.001$.} The determination of $c$ is sensitive to fitting of the data points and the initial transients should be excluded \cite{carlon}. {Therefore,} a careful determination of the TCP would require longer lengths and better statistics or equivalently longer CPU time.

Thermodynamic observables usually studied to characterise the nature of a denaturation transition are the number of bond pairs in contact per monomer ($n_c$), which also serves as the order parameter for the transition and the thermal response function ($C_c$) which is related to the fluctuation of $n_c$ and gives the specific heat after scaling with $T^2$; see Fig~\ref{fig:4}(a). Near, the transition $f\approx f_{sc}$ for a chain length $N$, we have the following scaling form for the specific heat \cite{vanderzande}
\begin{equation}
    C_c\sim N^{2\phi-1}g[(f-f_{sc})N^{\phi}],
    \label{eq:12}
\end{equation}
where $\phi$ is the crossover exponent, which determines the nature of the transition,  and $g$ is a scaling function. However, for large $N$ and close to $f_{sc}$ Eq.~\ref{eq:12} reduces to $C_{c,peak}\sim N^{2\phi-1}$. {We provide the data-collapse plot in Fig.~\ref{fig:4}(b), and the scaling of the specific heat peaks with the system size in Fig.~\ref{fig:4}(b) inset}. Our estimate of the crossover exponent $\phi=0.77\pm 0.001$ {for the zipping transition at $T=1.5$} compares well with an independent estimate using the bubble-size-distribution Eq.~\ref{eq:14} in the upcoming {paragraph}.

Although, we are interested in the thermodynamic limit ($N\rightarrow \infty$), study of these scaling laws for finite size systems near the critical point are often useful since single molecule DNA experiments are performed with finite size systems. These, critical points are characterized  by mathematical singularities of thermodynamical quantities which appear only for an infinite system, but are smeared for finite systems. This smearing is quantified by the smearing exponent which measures the rounding of the response curves near the transition point. It was shown that when a non-ordering field (other than temperature) drives the transition from first-order to second-order \cite{wortis,imry}, then the broadening near the transition, under the assumption that the broadening in the non-ordering field $(\mu)$ is the same as that of temperature gives that 
\begin{equation}
\Delta \mu_c \sim N^{-1/(2-\alpha)}.
\label{eq:13}
\end{equation}	  
where $\alpha=\left(2\phi-1\right)/\phi$ is the critical exponent for the divergence of specific heat and $1/(2-\alpha)$ is the smearing exponent. Although, the stretching force $f_s$, in our case, like temperature induces a transition, but since it is not coupled to the order parameter $(n_c)$ directly, it cannot be linked to any `ordering' field. Taking the width of the specific heat curves at the half-maximum as the measure of $\Delta f_s$ we plot the width of the $C_c$ curves for different sizes and then fitted with Eq.~\ref{eq:13}; see Fig.~\ref{fig:4}(c). We obtained $\alpha=0.72\pm 0.009$ and therefore $\phi\simeq 0.78$, compatible with the previously obtained result. The smearing exponent remains the same as that of the second-order smearing in temperature $\Delta T_c \sim N^{-1/(2-\alpha)}$. Thus, stretching a dsDNA provides an excellent system of studying this smearing exponent for a field which induces a transition to a more orderly state (fully stretched and bound) although not connected directly to the order parameter while also changing the nature of the transition to {second}-order.

{\it Bubble statistics:} In our model, a bubble is identified to be a {continuous} section of broken bonds flanked by the bound segments {on either side}. Thus, the broken bonds in the Y-fork do not count as part of the bubbles; see Fig.~\ref{fig:1}(d). 
In the zero-force limit $(f_s= 0)$ i.e. in the linear response regime, the isotropic {e.r.f} $\bar{\kappa}_{\rm cm}$ is controlled by 
these bubbles for a continuous transition and  by the broken bonds in the Y-fork region for a first-order transition \cite{dm}. Naturally, we expect some new behavior in the system under a finite force $(f_s\neq 0)$ when the fluctuations transverse to the direction of the applied force gets suppressed, since these thermal fluctuations are essential in determining the elasticity of the bulk DNA in the form of bubbles and Y-fork. 

In the process of bringing the strands together, the stretching force primarily aids in the formation of bubbles, the average bubble length ($l_b$) grows with the stretching force, peaking near the transition point and then dies out as the strands gradually collapse into a single strand; see Fig.~\ref{fig:5}(b). Averaging is done over length and then over configurations. Bubbles under a stretching force grow larger on an average in comparison to the thermal melting transition \cite{dm}. This is due to the combined effect of strong thermal fluctuations (at $T>T_m$) and the ordering force $f_s$. Note, that the peaks for average number of bubbles per monomer $(n_b)$ and the average length of bubbles $(l_b)$ shifted on either side of the transition point; see Fig.~\ref{fig:5}(a) and (b). This indicates that as the chain approaches the transition point from below $f_s\rightarrow f_{sc}-$, a few larger bubbles break to form many smaller bubbles in the process of stretching. The fluctuation in the number of bubbles per monomer $(C_b)$ gets large near the transition point; see Fig.~\ref{fig:5}(a) inset. In Fig.~\ref{fig:5}(b) inset we plot the fraction of broken bonds comprising the bubbles ($f_b$) and the Y-fork ($f_Y$). The bubble-size-distribution (bsd)  near the transition point follows a power law scaling \cite{carlon1} 
\begin{equation}
    P(l,N)\sim l^{-c}g(l/N),
\label{eq:14}    
\end{equation}
where the exponent $c$ is related to the reunion exponent of random walkers \cite{smsmb} and determines the nature of the transition and $g(l/N)$ is a scaling function; see Fig.~\ref{fig:5}(c). Away from the transition point $(f>f_{sc})$ an exponential distribution $P(l,N)\sim \exp{(-l/l_0)}$ is followed. As mentioned  previously, for $c\geq 2$ the transition is first-order, for $1<c<2$ the transition is second order and for $c<1$ there is no transition at all. The bsd exponent $c$ is related to the crossover exponent $\phi$ as
\begin{equation}
    \phi=c-1,
\end{equation}{}
with $\phi=0.77$, estimated independently from the specific heat curves; see Fig.~\ref{fig:4}(b). The bsd for stretching induced renaturation transition at $T=1.5$ and $f_s=2.45$ {for chain length $N=500$} follows an exponent $c=1.79 \pm 0.01$; see Fig.~\ref{fig:5}(c). This further corroborates the continuous nature of the transition \cite{hanke}. Nature of the transition do not changes on including semiflexibility, as observed from the exponent $c$ for $\eta=1$ at $T=1.5$ and $f_s=1.5$. Although, the possibility that whether an increased stiffness can drive the transition towards first-order needs to be checked further.

\begin{figure}[t]
\centering
 \hspace*{-8cm}(a)\\\includegraphics{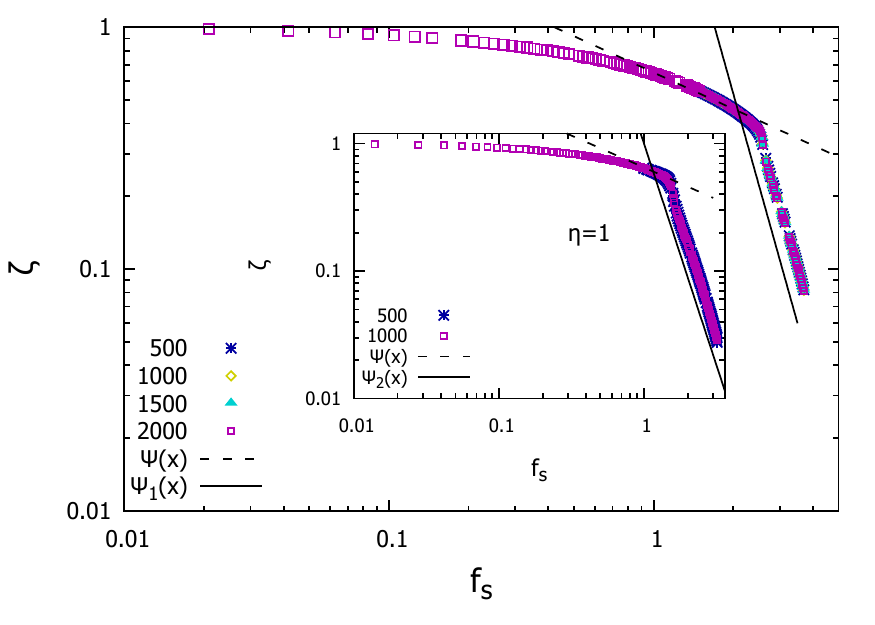}
 \hspace*{-8cm}(b)\\\includegraphics{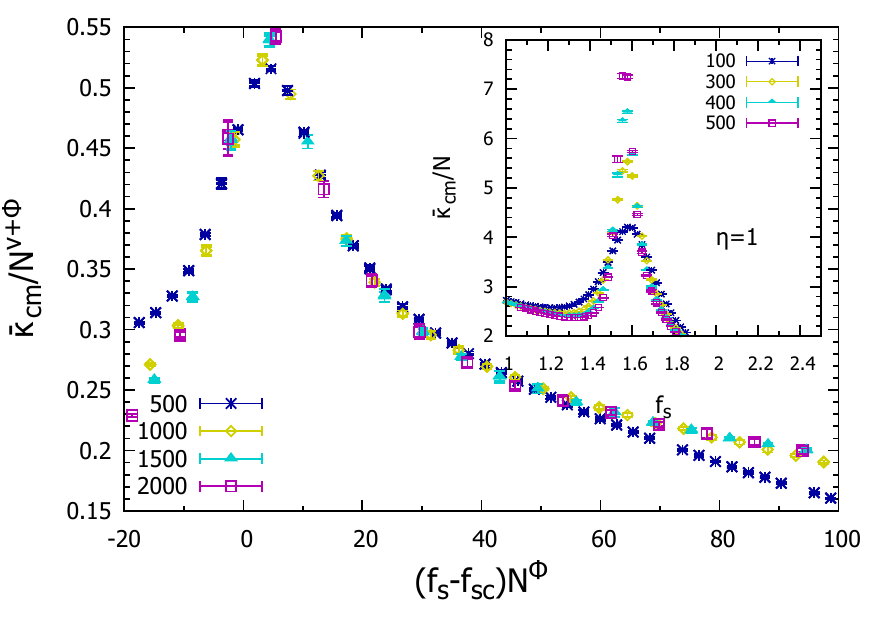}

\caption{(Color online) Average fractional extension and {e.r.f}. (a) Log-log plot of fractional extension $(\zeta)$ of strand $1$ along the direction of the stretching force ($f_s$), for a flexible DNA ($\eta=0$) at $T=1.5(>T_m)$ for chain lengths up to $N=2000$. The straight solid line represents $\Psi_1(x) \sim x^{-3.91}$. (Inset) Same as (a), but for a semiflexible DNA ($\eta=1$) at $T=1.5$ and chain lengths up to $N=500$. The straight solid line represents $\Psi_2(x) \sim x^{-3.48}$. For both cases the straight dotted line represents $\Psi(x)\sim x^{-0.5}$ represents WLC scaling. (b) Data collapse of the {e.r.f} $(\bar{\kappa}_{\rm cm})$ for a flexible DNA $\eta=0$ at $T=1.5$ for chain length up to $N=2000$ according to Eq.~\ref{eq:16} using $\nu=0.5$ and $\phi=0.77$. (Inset) e.r.f $\bar{\kappa}_{\rm cm}/N^{2\nu}$ with $\nu=0.5$ for a semiflexible DNA $\eta=1$ at $T=1.5$ and chain length up to $N=500$.}
    \label{fig:6}
\end{figure}

\begin{figure}[t]
\centering
\hspace*{-8cm}(a)\\\includegraphics{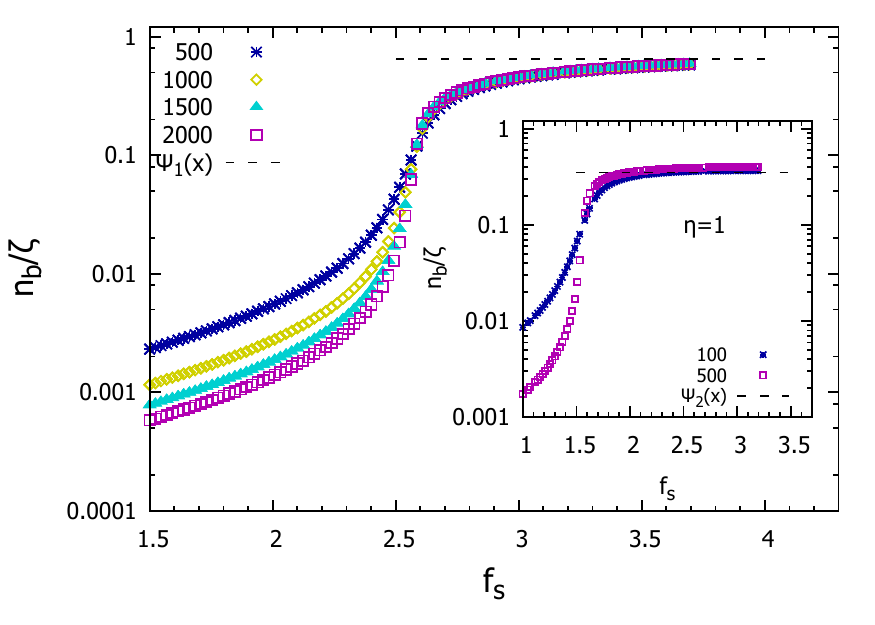}
\hspace*{-8cm}(b)\\\includegraphics{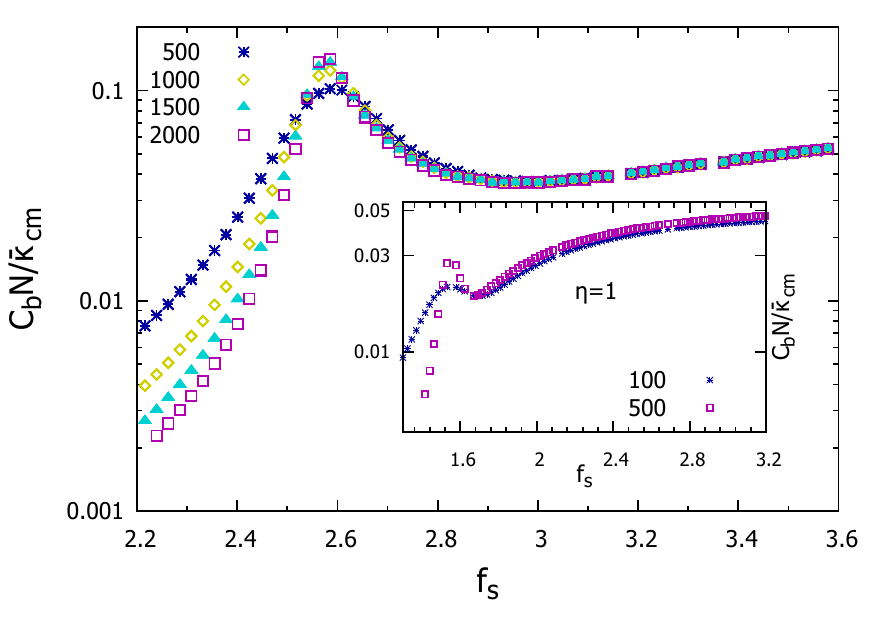}
\caption{(Color online) Relation between bubble related quantities and {e.r.f}. (a) Ratio of bubble number per monomer $(n_b)$ and the fractional extension $(\zeta)$ of strand 1 along the direction of force for flexible model $\eta=0$ at $T=1.5$ and chain length up to  $N=2000$. The {dotted} line $\Psi_1(x)=0.65$ is the value of the collapsed curve.  (Inset) Same as (a) but for semiflexible model $\eta=1$ and $T=1.5$ and chain length up to $N=500$. The {dotted} line $\Psi_2(x)=0.35$ represents the value of the collapsed curve. (b) Fluctuation in number of bubbles {per monomer} $(C_b)$ scaled by the {e.r.f} $(\bar{\kappa}_{cm}/N)$ for $\eta=0$ and $T=1.5$ and chain length up to $N=2000$. (Inset) Same as (b) but for a semiflexible chain with $\eta=1$ and $T=1.5$ for chain length up to $N=500$.}
\label{fig:7}
\end{figure}

{\it Elastic properties:} The elastic  properties for single polymer chains are studied using the force-extension curves where the slope of the curve provides an estimate of its extensibility. However, for systems like dsDNA inter-strand correlations plays important role in determining the bulk elasticity and the elastic response of individual strands do not capture the whole picture. Therefore, we study the fluctuation response of the c.m. chain to investigate the elastic behavior of the system.

In Fig.~\ref{fig:6}(a) and Fig.~\ref{fig:6}(b) we plot the fractional extension {$(\zeta)$} of {an individual} strand {along the force} and the {e.r.f} {per monomer ($\bar{\kappa}_{\rm cm}/N$)}  respectively for $\eta=0$ and $1$. The continuous variation of $\zeta$ around the transition point is the signature of a  second-order transition. Note that, post-transition, the extension happens faster, both for $\eta = 0$ and $1$ and follows a power law. The {e.r.f} $\bar{\kappa}_{\rm cm}$ {diverges} at the transition point, without any pre-transitional signature in the thermodynamic limit $N\xrightarrow{} \infty$; see Fig.~\ref{fig:6}(b). The peak for $\bar{\kappa}_{\rm cm}/N$ increases with the size of the DNA and eventually leads to a $\delta$-function in the infinite chain limit. Although, for finite size systems, the {e.r.f} per monomer is continuous and becomes length independent away from the transition on either side. This indicates strong finite size effects mainly around the transition point over a range of forces. Further, this {divergence} in $\bar{\kappa}_{\rm cm}$ grows sharper with the stiffness of the chains. A finite-size scaling should be of the form 

\begin{equation}\label{eq:16}
\bar{\kappa}_{\rm cm}/N^{2\nu}=N^{\phi-\nu}g[(f_s-f_{sc})N^{\phi}].
\end{equation}

The system undergoes a change in flexibility from the most flexible state when it is unbound with large $\bar{\kappa}_{\rm cm}$ and zero external force $f_s= 0$, to a rigid state with $\bar{\kappa}_{\rm cm}$ going {towards} zero. Although under stretching force the renaturation transition is {continuous at $T=1.5$}, the profile for $\bar{\kappa}_{\rm cm}$ is remarkably different from that of a DNA constructed with two Gaussian chains undergoing {a continuous} thermal melting transition where the $\bar{\kappa}_{\rm cm}$ follows the order parameter curve without any jumps near the transition point \cite{dm}.

Next, we investigate whether there is any connection between the observables describing the elastic response of the system {\it viz.} $\zeta$ and $\bar{\kappa}_{\rm cm}$, with the bubble related quantities such as $n_b$ and $C_b$, which controls the nature of the thermal transition \cite{pal}.  Since the  {e.r.f} $\bar{\kappa}_{\rm cm}$ is associated with the fluctuations in the average extension of the c.m. $\zeta_{\rm cm}$, we expect $\bar{\kappa}_{\rm cm}$ to be related to the fluctuation in bubbles $C_b$ while $\zeta_{\rm cm}$ or simply $\zeta$ should be related to $n_b$. In Fig.~\ref{fig:7}(a) and Fig.~\ref{fig:7}(b) we plot the quantities $n_b/\zeta$ and $C_bN/\bar{\kappa}_{cm}$. We found that beyond the transition point, the curves for the two ratios collapse into a single master curve which indicates that $\zeta\sim n_b$ and $\bar{\kappa}_{cm}/N\sim C_b$ with a weak dependence on $f_s$. {From this proportionality}, $n_b$ and $C_b$ seems to be an important element in determining the elastic response of the bound system containing thermally melted regions or bubbles under a stretching force. Moreover, only bubbles contribute to the flexibility of the stretched DNA in the partially bound state with finite fraction of broken bonds in the bubbles $f_b=n_b\times l_b\neq0$, while the fraction of broken bonds in the Y-fork ${\rm f_Y}\rightarrow{0}$; see Fig.~\ref{fig:5}(b) inset.

\begin{figure}[t]
    \centering
    \includegraphics{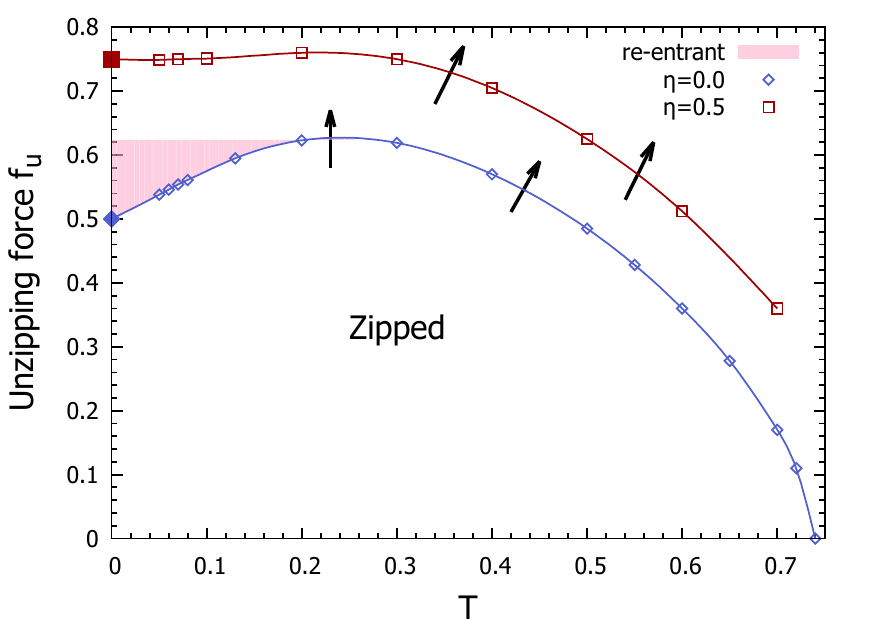}
    \caption{(Color online) Unzipping phase diagram of a flexible $(\eta=0)$ and semiflexible $(\eta=0.5)$ DNA. The smooth line is an interpolation using the data points. The arrows along the phase boundary directs towards the unbound phase. Low temperature slope is determined by the entropy of the bound phase.  As it approaches zero for {non-zero $\eta$}, curve becomes horizontal {as shown for $\eta=0.5$ (open squares)}. Analytical values of $f_{uc}$ at $T=0$ are in exact match with the simulation results both for $\eta=0$ and $0.5$.}
    \label{fig:8}
\end{figure}

\section{Elastic Properties under an Unzipping Force}										
\textit{The unzipping transition:} A DNA can be mechanically unzipped \cite{smb1} by applying an equal and opposite force $f_u$ on the two open ends of the DNA; see Fig.~\ref{fig:2}. Unlike, a stretching force, an unzipping force tries to separate the DNA into two single strands. The unzipping takes place only after the force exceeds a critical value $f_{uc}$ \cite{smb1}. This critical value could depend upon factors like the temperature, flexibility etc. While theoretical studies have obtained the force-temperature phase diagram \cite{smb2}, but the agreement with the experimental curve for the unzipping of a {\it lambda phage} DNA is only over a selected range of temperature outside which it differs significantly, resulting into either under-estimation or over-estimation of the critical force \cite{dani1}. One of the major factors that was not taken into consideration in these previous theoretical and simulative investigations in studying the unzipping phase diagram in 3d, is the large difference in the rigidity between the ss and ds segments. Here, we perform a simulative study, of the effect of semiflexibility of the ds segments of the DNA on the unzipping phase diagram along with the elastic properties and bubble statistics, for unzipping induced by an externally applied force. 

\begin{figure}[t]
    \centering
     \hspace*{-8cm}(a)\\\includegraphics{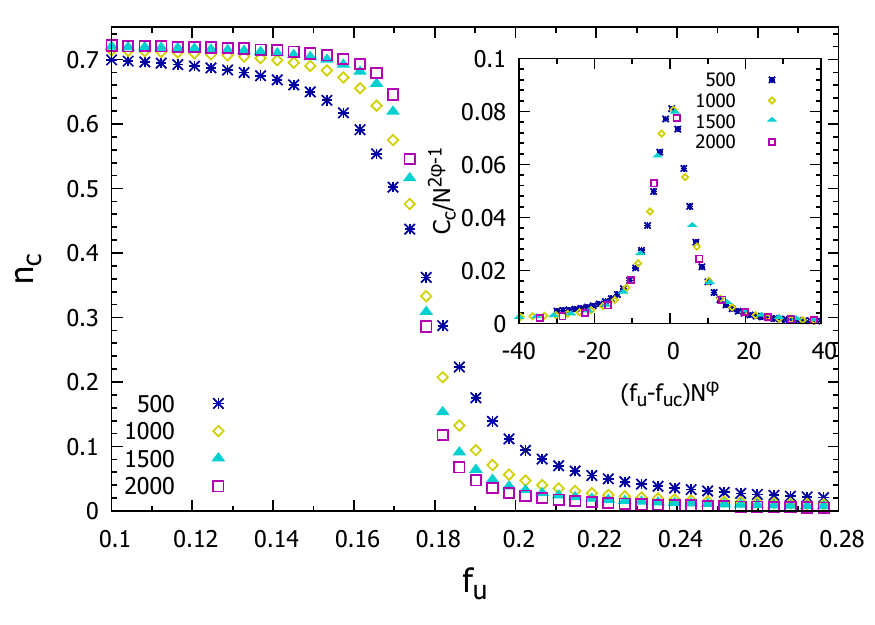}
     \hspace*{-8cm}(b)\\\includegraphics{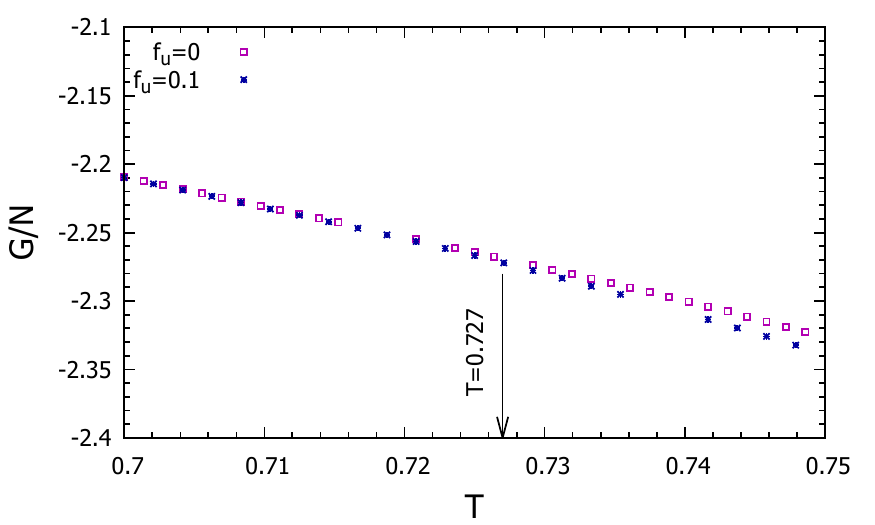}
    \caption{(Color online) Order parameter, its fluctuation and Gibbs free energy. (a) Number of base-pair contacts per monomer $(n_c)$. (Inset) {data collapse for contact number fluctuation $(C_c)$ per monomer, using $\phi=0.96\pm 0.02$ and $f_{uc}=0.176(7)$}, for a flexible DNA $(\eta=0)$ {subjected to} unzipping force $(f_u)$ at a constant temperature $T=0.7(<T_m)$ for chain lengths upto $N=2000$. (b) Comparison of {the two} free energies per monomer $(\mathcal{G}/N)$ for $f_u=0$ and $f_u=0.1$ for chain length $N=400$. Arrow directs to the point of difference between the two free energies representing $T_c$ for the $f_u=0.1$ unzipping force; see Fig.~\ref{fig:8}.}
    \label{fig:9}
\end{figure}
The force-temperature phase diagram of a DNA, unzipped through pulling of both the strands simultaneously in opposite directions, consisting of flexible chains, show a low-temperature denaturation or re-entrant phase transition due to the non-zero ground state entropy \cite{smb_mdz}; see Fig.~\ref{fig:8}. {Unzipping  of a flexible DNA ($\eta=0$) at $T=0.7$ is found to be weakly first-order with $\phi=0.96\pm0.02$ and $f_{uc}=0.176(7)$ obtained from the data collapse of specific heat; see Fig.~\ref{fig:9}(a) inset. The specific heat peaks scales roughly $\propto N$. Since our maximum length is only upto $N=2000$ this might introduce corrections to scaling. Our results are in agreement with the thermal melting of a DNA where $\phi=0.98\pm 0.15$ is estimated from the scaling of peaks of the specific heat with length \cite{causo,comm5}. This shows that under an unzipping force the nature of the transition remains intact. Further, the nature of the transition do not changes by semiflexibility of the ds segments or by the magnitude of the unzipping force.} The only effect of the semiflexibility is to induce stability into the system by lowering the entropy of the bound phase {and thereby the formation of bubbles \cite{dm}}. This also affects the low-temperature denaturation transition, since the re-entrant phase transition is solely driven by the entropy of the ground state. Therefore, for semiflexible chains, the low-temperature denaturated phase vanish; see Fig.~\ref{fig:8}.

A simple scaling argument for the critical force near the thermal melting  follows from a thermodynamical analysis of unzipping for a {\rm Y}-model as \cite{smb2}
 \begin{equation}
     f_{uc}(T)\sim T^{1-\nu} \left(1-\frac{T}{T_m}\right)^{\nu},
\label{eq:17}     
 \end{equation}
where $T_m$ is the thermal melting point and $\nu$ is the size exponent. That the scaling near the thermal melting point can be explained by a simple {\rm Y}-model of DNA is due to the fact that the unzipping force is unaware of the bubbles residing beyond the {\rm Y}-fork.  Near the low temperature unzipping transition for $\eta=0$ \cite{smb2},
 \begin{equation}
     f_{uc}(T)=\frac{1}{2}(\epsilon+T\log\mu_z)
 \label{eq:18}
 \end{equation}
where $\mu_z$ is the effective coordination number of the DNA. Thus, the curvature of the re-entrant transition $\left(\frac{\partial f_{uc}}{\partial T}\right)_{T\rightarrow 0}$ is controlled by the entropy (or in other words the flexibility) of the ds bound phase. Thus, setting $T=0$ we obtain the zero-temperature unzipping force $f_{uc}(T=0)=0.5$. Similarly, for a semiflexible DNA ($\eta \neq 0$), Eq.~\ref{eq:18} can be written as $f_{uc}(T=0)=(\epsilon+\eta)/2$, giving the zero-temperature unzipping force for $\eta=0.5$ to be $f_{uc}(T=0)=0.75$  in exact match with the {extrapolation of the} simulation results and the slope $\left(\frac{\partial f_{uc}}{\partial T}\right)_{T\rightarrow 0} \rightarrow 0$ with the ground state entropy going to zero; see Fig.~\ref{fig:8}. Thus, resulting in the vanishing of the re-entrant transition {for non-zero $\eta$}. Estimate of the transition points are obtained from the peaks of the specific heat curves; see Fig.~\ref{fig:9} \cite{comm2}.

\begin{figure}[H]
    \centering
     \hspace*{-8cm}(a)\\\includegraphics{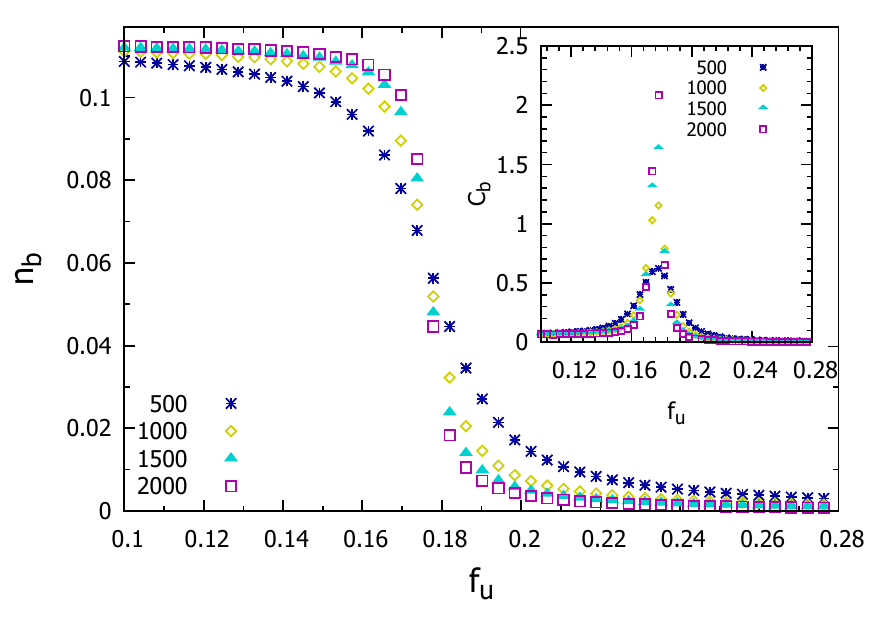}
     \hspace*{-8cm}(b)\\\includegraphics{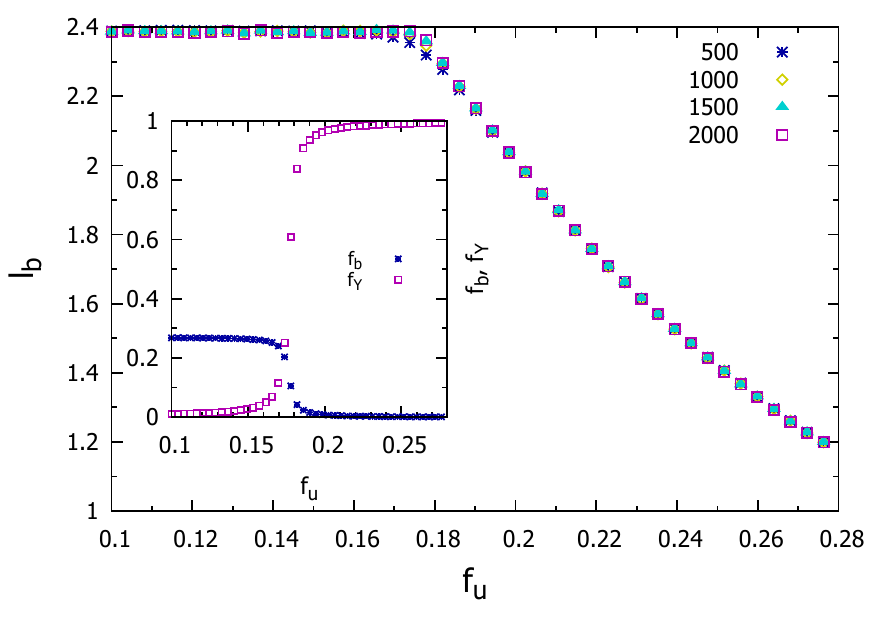}
    \caption{(Color online) Bubble number, its fluctuation and average bubble length of a DNA under unzipping force. (a) Average number of bubbles per monomer $(n_b)$. (Inset) Bubble number fluctuation per monomer $(C_b)$. (b) Average length of bubbles $(l_b)$ along the chain. Average is taken along the chain and then over configurations. (Inset) Fraction of broken bonds $(f_b)$ forming bubbles and Y-fork $(f_Y)$. All data taken for a flexible DNA i.e. $\eta=0$ at $T=0.7(<T_m)$.}
    \label{fig:10}
\end{figure}
{\it Hypothesis of the impenetrability of force:} From a thermodynamical viewpoint we have two mutually exclusive situations, where either the force ($f_u$) or the extension ($z$)  is fixed.  These two scenarios correspond  to the two possible ensembles in the statistical mechanical picture.  The fixed-{distance} ensemble is characterised by the Helmholtz free energy $\mathcal{F}(T,z)$ and the fixed-{force} ensemble is characterised by the Gibbs free energy $\mathcal{G}(T,f_u)$. For a system which is both thermally and mechanically coupled to the environment, we need to consider the change in the Gibbs free energy.  The mechanical coupling comes from the applied force at the endpoints {and the two free energies are related} via a Legendre transformation
\begin{equation}
    \mathcal{G}(T,f_u)=\mathcal{F}(T,z)-f_uz,
\label{eq:19}    
\end{equation}
where $f_u$ is the force and $z$ is the extension of the strand along the direction of the applied force. For a first-order unzipping transition $\mathcal{G}(T,f_u)$ is continuous across the phase boundary which implies $\mathcal{G}_{z}(T,f_{uc})=\mathcal{G}_{u}(T,f_{uc})$, where $\mathcal{G}_z  (\mathcal{G}_u)$ represents the free energies in the zipped(unzipped) phase. Hypothesizing that the force do not penetrate the bound state for $f_u<f_{uc}(T)$ we can write \cite{sadhukan}
\begin{equation}
    \mathcal{G}_z(T,f_u)=\mathcal{G}_z(T,0),~~~(f_u\leq f_{uc}),
\label{eq:20}    
\end{equation}
i.e. the Gibbs free energy in the presence of the unzipping force must be equal to the Gibbs free energy in absence of the unzipping force in the zipped phase.  We estimate the free energy ($=-\beta^{-1}\ln{\mathcal{Z}}$) where $\mathcal{Z}$ is the {canonical} partition function estimate {in the fixed force ensemble} directly from the \textsc{PERM} simulations and plot the free energies for $f_u=0$ and $f_u=0.1$ in Fig.~\ref{fig:9}(b).  The point of difference between the two free energies then should give the critical point of unzipping.  This hypothesis is valid irrespective of the stiffness of the chain. Although a thermodynamical description of the first-order unzipping transition rests on the hypothesis of the {\it impenetrability of force}, thermodynamics does not rule out the possibility of a continuous unzipping transition in case there are sources, that allow for the force to penetrate \cite{sadhukan,sadhukan1}.

\begin{figure}[b]
    \centering
    \includegraphics{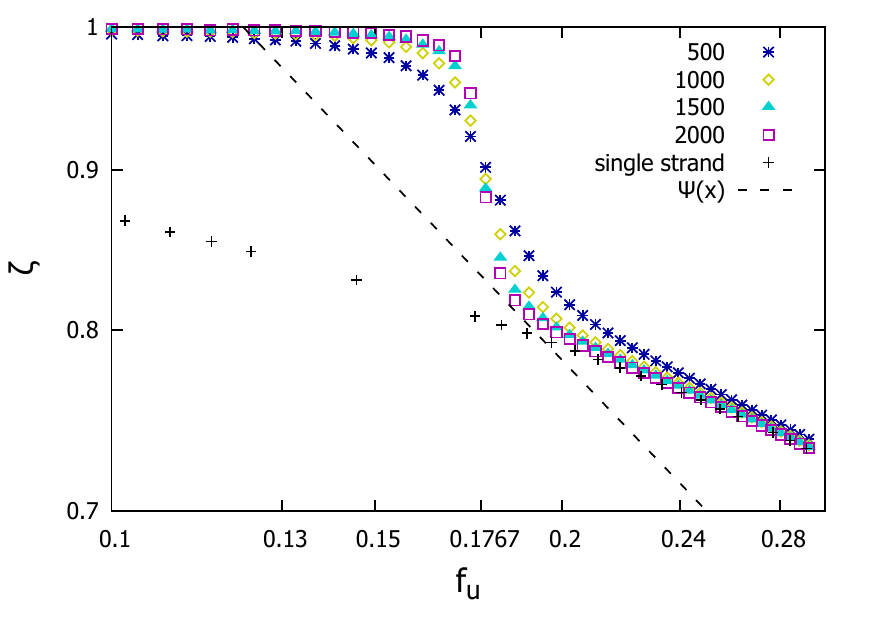}
        \caption{(Color online) Fractional extension ($\zeta$) of strand 1 along the direction of the unzipping force. The straight dotted line represents $\Psi(x)\sim x^{-0.5}$ while single strand refers to the extension along the force for a single {flexible} polymer chain.}
    \label{fig:11}
\end{figure}
{\it Bubble statistics:} {The important length scales of the problem come from the bubble size along the chain $(\tau)$ and in spatial extent $(\xi)$ \cite{smb1}. $\tau$ is related to the average bubble length ($l_b$). In Fig.~\ref{fig:10}(a) we plot the average number of bubbles per monomer $(n_b)$, in Fig.~\ref{fig:10}(b) the average bubble length $(l_b)$ and (inset) the fraction of broken bonds forming the bubbles $(f_b)$ and the Y-fork $(f_Y)$ for unzipping at $T=0.7$.} The value of $n_b$ {in the bound phase} represents the corresponding zero-force value at $T=0.7$. The impenetrability of the unzipping force below a certain critical value, make the bubbles, residing deep within the chain, impervious to the external force.  The average number of bubbles per monomer remains constant up to the critical point $f\rightarrow f_{uc}^-$. Similarly, $l_b$ do not changes until the critical point is reached. {Further, in finite-size systems a Y-fork may coexist with finite fraction of broken bonds in the bubbles near the critical point; see Fig.~\ref{fig:10}(b) inset.} Since, the bubbles remain invisible to the external force which acts only as a boundary effect, the nature of the unzipping transition remains intact.

\begin{figure}[t]
    \centering
     \hspace*{-8cm}(a)\\\includegraphics{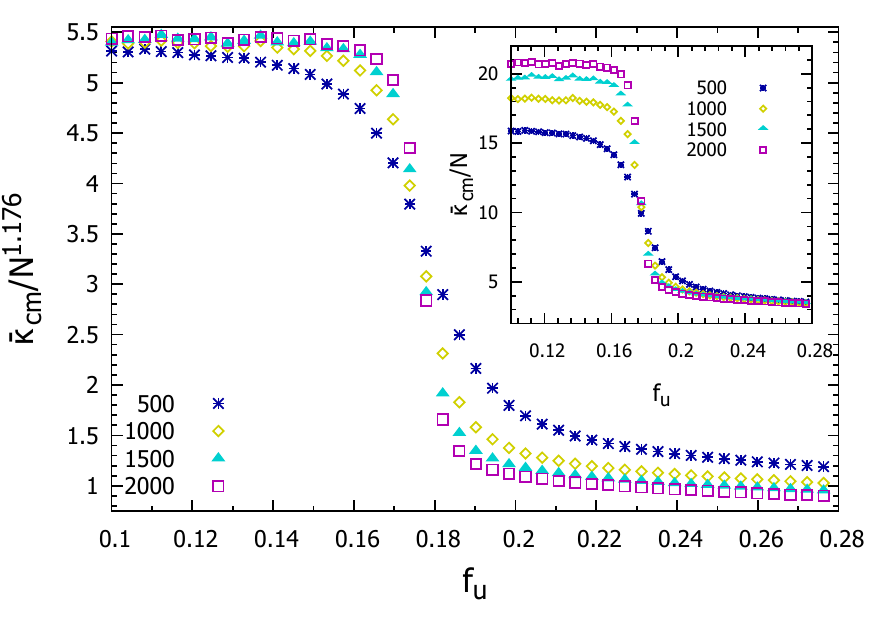}
     \hspace*{-8cm}(b)\\\includegraphics{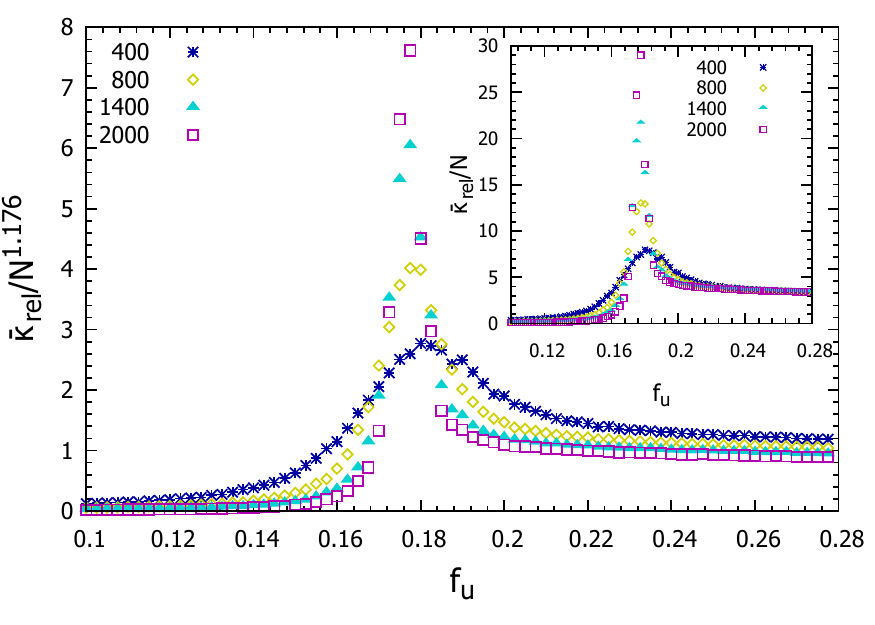}
        \caption{(Color online) (a) {Rescaled {e.r.f} from c.m. chain $(\bar{\kappa}_{\rm cm}/N^{1.176})$ against unzipping force $(f_u)$. (Inset) same as (a) but when scaled by $N$. (b) Rescaled {e.r.f} from rel chain $(\bar{\kappa}_{\rm rel}/N^{1.176})$ against unzipping force $f_u$. (Inset) same as (b) but scaled by $N$. All data taken for a flexible DNA ($\eta=0$) at $T=0.7(<T_m)$. Notice that the value in the zipped phase remains the same as the zero-force value \cite{dm} while a reduction took place in the unzipped phase.}}
    \label{fig:12}
\end{figure}
{\it Elastic properties:} We plot the average fractional extension $\zeta$ of an individual strand along the force in Fig.~\ref{fig:11} and the isotropic {e.r.f} obtained from the c.m. chain $\bar{\kappa}_{\rm cm}$ and {from the} relative chain $\bar{\kappa}_{\rm rel}$ in Fig.~\ref{fig:12}. The sharp drop in $\zeta$ signals a first-order transition; see Fig~\ref{fig:11}(a). That, the extension remains zero till the critical point is reached is another instance of {the} {\it impenetrability of force}. In the large force limit $(f\gg f_{uc})$ the extension $(\zeta)$ follows the curve of the single strand. The qualitative behavior of the {e.r.f} obtained from the {\rm c.m.} chain $\bar{\kappa}_{\rm cm}$ is similar to {that of} the thermal melting {\cite{dm}}, while that obtained from the {\rm rel} chain is {divergent} at the transition point {; see Fig.~\ref{fig:12}(b). This divergence is absent in thermal melting \cite{dm}}. It is evident from Fig.~\ref{fig:12} that even in the presence of an external force the isotropic part of the {e.r.f} $\bar{\kappa}_{\rm cm}$ ({or $\bar{\kappa}_{\rm rel}$}) in the bound phase represents the zero-force value {with a scaling exponent of $\nu=0.588$}. While in the unzipped phase a reduction took place w.r.t the zero-force value {when scaled by $N^{1.176}$}, owing to the presence of the unzipping force. {Although, a better collapse is obtained when $\bar{\kappa}$ is scaled by $N$ with $\nu=0.5$}. This shows that given the bubbles remaining unaffected by any boundary effect, bulk elasticity do not changes. On the contrary, in the stretching case since bubbles gets modified continuously, this leads to a gradual {change} in the value of $\bar{\kappa}_{\rm cm}$ starting from the zero-force value {and also after the transition}. {This is evident from the plots of the average bubble length and average number of bubbles; compare Fig.~\ref{fig:5}(b) and Fig.~\ref{fig:10}(b)}. Thus, {the} \textit{impenetrability of force} \cite{sadhukan} plays an important role in controlling the elastic properties of the zipped phase. We expect this behavior independent of the position where the force is applied. Also noticeable is that $\bar{\kappa}_{\rm cm}$ and $\bar{\kappa}_{\rm rel}$ behaves in the similar way as the stiffness constants of the restoring potential for fluctuations about the mean value of the order parameter along the longitudinal ($\kappa_l$) and transverse ($\kappa_t$) directions respectively, derived from the Landau-Ginzburg Hamiltonian \cite{kardar} i.e. $\bar{\kappa}_{\rm cm}(\kappa_l)=\bar{\kappa}_{\rm rel}(\kappa_t)$ in the unbound (disordered) phase and $\bar{\kappa}_{\rm rel}(\kappa_t)=0$ in the bound (ordered) phase. This vanishing of the transverse component $(\kappa_t=0)$ represents the Goldstone modes appearing due to spontaneous breaking of a continuous symmetry. Although, such a symmetry breaking is not known for DNA. 

\section{Conclusion}
To conclude, we consider a minimal model of a dsDNA in good solvent conditions, to study {the change in} the elastic {response} under {a change in the state of the system when subjected to} unzipping or stretching forces and investigate how the elements that contribute to the flexibility {\it viz.} number of bubbles and its fluctuation gets modified by it. Interestingly, the elastic response of the DNA is different for the two types of forces considered. The elasticity for the flexible case is completely entropic, which emerges due to the bond-bond correlation. On the other hand, the semiflexible model contains an additional intrinsic rigidity. We found that a stretching force alters the bubble statistics, and hence the order of the transition resulting in a continuous transition, {for sufficiently strong forces}, even with excluded volume interaction. With semiflexibility in the ds segments, the DNA is found to cooperate with the stretching force in the sense that the renaturation transition occurs at a lower critical force in comparison to the flexible case. We also show that the average extension along the force and the {e.r.f} is related to the average bubble number and its fluctuation respectively and that the flexibility of the bound phase is mainly due to the bubbles because of the complete disappearance of the Y-fork. 

On the other hand, for an unzipping force, the unzipping transition remains first-order as in the case of thermal melting for all values of forces and semiflexibility. Unzipping takes place at a higher critical force for systems at lower temperature. Semiflexibility provides thermal {and mechanical} stability against an unzipping force.  Although, our results remain more towards the theoretical models but it surely narrows the different possibilities e.g. semiflexibility do not seems to change the nature of the flexible chain phase diagram except the vanishing of the low temp re-entrant part. We also revisited some previously known results from the perspective of the model we describe here. Additionally, we show that a modification of the bubble states are necessary in order to change the bulk elasticity of the DNA. {Intriguingly, the elastic response is largest at the critical point. This seems to relate to the notion that biological systems poise itself at the criticality for enhanced elastic response \cite{mora}.}

For simplicity, we chose equal forces for both the strands. Although, two unequal forces in arbitrary angles, can be decomposed into stretching and unzipping {forces} applied at the end independently. This allows for an easy experimental implementation. Other interesting situations include  position dependent elastic response of the DNA, which might show new features different from the simple minded stretching at the endpoints e.g. how critical force for unzipping depends upon the position where the force is being applied \cite{kapri1,kapri2}, the heterogeneity of the base-pair sequence or elastic response under spatial confinement etc. We hope our results will serve as a theme for future experiments on DNA. 

\section{Acknowledgement}
D.M. thanks Somendra M Bhattacharjee for insightful discussions. The computer simulations were performed on the SAMKHYA high-performance computing facility at the {\it Institute Of Physics}, Bhubaneswar.

{}

\begin{thebibliography}{99}
\bibitem{comm6} {Like a ribbon, a DNA is also associated with two elastic constants, one for twist and another for bending.}
\bibitem{comm3} {Assuming there is shape anisotropy \cite{haber} when generating the polymer chain, since the anisotropy is present isotropically in all directions during growth in different configurations, we should be able to take the trace.}
\bibitem{comm1} The number of free sites facing the end of a chain is the atmosphere of that step.
\bibitem{comm5} {Although a different estimate of $\phi=1.31(14)$ is obtained when calculated from fitting the shift in the position of the specific heat peaks \cite{causo}. This is close to the $\phi$ as obtained from the  bubble-size-distribution in \cite{dm}.}
\bibitem{comm2} At low temperatures, overflowing of the partition function estimation can be a major computational hindrance. This can be overcome by simply multiplying a constant factor ($<<1$) at each step and by using 128-bit quadruple precision floating point format variables.
\end{thebibliography}
\end{document}